\documentclass[aps,jcp,twocolumn,superscriptaddress,floatfix,10pt]{revtex4}

\usepackage{graphicx}
\usepackage{amsmath}
\usepackage{color}
\usepackage{xcolor}
\usepackage{dcolumn}
\usepackage{soul}
\usepackage{comment}
\usepackage{appendix}
\usepackage{hyperref}
\usepackage[capitalize]{cleveref}
\usepackage[thinc]{esdiff}

\begin{document}

\title{Computational indentation in highly cross-linked polymer networks}

\author{Manoj Kumar Maurya}
\affiliation{Department of Mechanical Engineering, Indian Institute of Technology Kanpur, Kanpur UP 208016 India}
\author{C\'eline Ruscher}
\affiliation{Department of Mechanical Engineering, University of British Columbia, Vancouver BC V6T 1Z4, Canada}
\author{Debashish Mukherji}
\email[]{debashish.mukherji@ubc.ca}
\affiliation{Quantum Matter Institute, University of British Columbia, Vancouver BC V6T 1Z4, Canada}
\author{Manjesh Kumar Singh}
\email[]{manjesh@iitk.ac.in}
\affiliation{Department of Mechanical Engineering, Indian Institute of Technology Kanpur, Kanpur UP 208016 India}

\date{\today}

\begin{abstract}
Indentation is a common experimental technique to study the mechanics of polymeric materials. 
The main advantage of using indentation is because this provides a direct correlation between the 
microstructure and the small-scale mechanical response, which is otherwise difficult within the 
standard tensile testing. Here, majority of studies have investigated hydrogels, microgels, elastomers, and even soft biomaterials. However, a lesser investigated system is the indentation in highly cross-linked polymer (HCP) networks, 
where the complex network structure plays a key role in dictating their physical properties. 
In this work, we investigate the structure-property relationship in HCP networks using the computational 
indentation of a generic model. We establish a correlation between the local bond breaking, the network rearrangement,
and the small-scale mechanics. The results are compared with the elastic-plastic deformation model. 
HCPs harden upon indentation.
\end{abstract}

\maketitle

\section{Introduction}
 Polymers are  ubiquitous since the early start of life in the form of natural rubber, starch, and cellulose~\cite{fang2005development, hon2017cellulose, chen2007acetylated, james1943theory}.
 However, the concept of synthetic polymers is relatively new, which was first proposed in the pioneering work of 
 Hermann Staudinger~\cite{staudinger1920}. Polymers are of particular interest because the relevant energy scale in these 
 materials is about $k_{\rm B}T$ at the ambient temperature, thus their properties are dictated by the large conformational 
 and compositional fluctuations \cite{kroger2004simple, muller2020process, DMCMMKK2020,singh2020glass}. 
 Here, $k_{\rm B}$ is the Boltzmann constant. 
 Therefore, polymers provide a suitable platform for the flexible design of advanced soft materials. For example, polymers are 
 widely used for the lubrication~\cite{klein1994reduction, de2014solvent, singh2015polymer,singh2020polymer}, the confinement mechanics of biological materials ~\cite{missirlis2014combined, kim2002afm, wen2014interplay, beamish2010effects}, the smart materials ~\cite{brighenti2020smart, stuart2010emerging,DMCMMKK2020}, thermoelectrics ~\cite{tripathi2020optimization, shi2017tuning}, and 
 in the common daily-use materials~\cite{kim2015high, halek1988relationship, jain2011biodegradable,maier2001polymers}, to name a few.

Traditionally linear polymers are commonly used for various applications \cite{kroger2004simple, muller2020process, DMCMMKK2020,kim2015high}, while the more recent interest have been directed 
towards the cross-linked polymers that range from elastomers (weakly cross-linked) to epoxies (highly cross-linked) ~\cite{stevens2001interfacial,MukherjiPRE2008, lv2021effect}. 
Cross-linked, in particular highly cross-lined polymers (HCP), are important because they are light-weight, high-strength 
materials that can also have self healing properties \cite{NancySotos,sharifi2014toughened,gold2017microscopic,Melissa2019}.   
Here, one of the important physical properties of HCPs, and polymers in general, is their 
mechanical response~\cite{stevens2001interfacial, MukherjiPRE2008, lv2021effect}. Therefore, a more in-depth 
understanding of the mechanics is needed for the advanced applications of polymeric materials with tunable
properties.

Experimentally, two most common techniques are the tensile (or shear) deformation \cite{NancySotos, sharifi2014toughened,rahil2016nanoscale, persson2018some}  
and the nano-indentation~\cite{mathis2019indenting, mathis2018two, singh2018combined, muser2019modeling, akhtar2018oscillatory, kalcioglu2012macro}. The former technique has been extensively employed in the 
standard experimental setups \cite{NancySotos,sharifi2014toughened,lv2021effect} and also in a variety of computational studies~\cite{stevens2001interfacial, MukherjiPRE2008}, which gives the 
bulk mechanics. However, within the tensile studies, the localized force response due to 
the small-scale complex structures becomes exceedingly difficult to extract. 
In this context, mechanical indentation may serve as a better technique, where extensive recent experimental efforts 
have been performed \cite{mathis2018two,mathis2019indenting,backes2017combined}, while the computational indentation studies in polymers are rather limited \cite{boots2022quantifying}. 
Additionally, the indentation experiments are used to measure the hardness of materials and have the 
basic goal of quantifying materials' resistance to plastic deformation.

Indentation-based techniques have been extensively employed for the polymeric systems, such as hydrogels, microgels, elastomers~\cite{singh2018combined, muser2019modeling, mathesan2016molecular} and also often used to measure the mechanical stiffness is the biological cells~\cite{nanoindentationEbenstein2006,efremov2019anisotropy}. Moreover, the similar studies in HCP are limited.
Here, using large scale molecular dynamics simulations of a generic model, we have studied the 
mechanics of HCP networks with different functionalities using computational indentation. 
To the best of our knowledge, until now the computational works have ``only" employed the tensile deformation
for polymers ~\cite{harmandaris2000atomistic, kroger1997polymer, murashima2021viscosity, kim2014plastic, aoyagi2000molecular, parisi2021macro} and the indentation simulations are usually performed on the crystalline nanostructures~\cite{chen2018nanoindentation}.
Therefore, the investigation discussed herein is a first attempt that employs the indentation technique 
for HCPs within a generic mesoscale framework.
 
We note in passing that for this study we have employed a (chemically independent) generic model. While generic models 
are extremely useful to make qualitative comparisons with the experiments, they do not provide any quantitative agreement 
with a chemical specific system. Moreover, for the HCP networks, very little experimental information is 
known regarding the network microstructures and thus even an all-atom representation of HCP 
does not guarantee any realistic case known from a native chemical system. 
Therefore, we have deliberately chosen to use generic model where tuning 
the system parameters for the desired property is rather trivial.

The remainder of the paper is organized as follows: In Section \ref{sec:method}, we sketch our methodology. 
Results and discussions are presented in Section \ref{sec:res}, and finally, the conclusions are
drawn in Section \ref{sec:disc}.

\section{Model and method}
\label{sec:method}

HCP networks with two different network functionality $n$ are chosen for this study. Here, $n$ defines 
the maximum number of bonds that a monomer can form with its neighboring monomers. 
We have chosen a system of tri-functional (i.e., $n = 3$) and a tetra-functional (i.e., $n = 4$) networks. 
The systems consist of $N = 2.53 \times 10^5$ LJ particles randomly distributed within a 
cubic box at an initial monomer number density $\rho_{\rm m} = 0.85\sigma^{-3}$. 
The simulations are performed using the LAMMPS molecular dynamics package~\cite{plimpton1995fast}.

\subsection{Interaction potentials}

We employ a generic molecular dynamics simulation approach. Here, the non-bonded
monomers interact with a 6$-$12 Lennard-Jones (LJ) potential 
$u_{\rm nb}(r) = 4\epsilon \left[\left({\sigma}/{r}\right)^{12}-\left({\sigma}/{r}\right)^{6}\right]-u_{\rm cut}$ 
if the distance $r$ between two monomers is less than a cutoff distance $r_c = 2.5\sigma$. $u_{\rm non-bonded}(r) = 0$ for $r > r_c$, 
and $u_{\rm cut}$ is chosen such that the potential is continuous at $r_c$.
Here, $\epsilon$ and $\sigma$ are the LJ energy and the LJ length, respectively. 
This leads to a unit of time $\tau = \sigma \sqrt{m/\epsilon}$, with $m$ being the mass of the monomers.
The values representative of the hydrocarbons are, $\epsilon = 30$ meV, $\sigma=0.5$ nm, and $\tau = 3$ ps. The
unit of pressure $P_{\circ} = 40$ MPa~\cite{kremer1990dynamics}.

The equations of motion are integrated using the velocity Verlet algorithm with a time step $0.005\tau$ and the 
temperature is set to $T = 1\epsilon/k_{\rm B}$ that is much higher than the typical glass transition temperatures $T_{\rm g}\simeq 0.4\epsilon/k_{\rm B}$ in these network systems~\cite{stevens2001interfacial,MukherjiPRE2008}, thus representing a HCP gel phase. The temperature is imposed using a Langevin thermostat with a 
damping coefficient of $\gamma = 1\tau^{-1}$. The initial LJ system is equilibrated for $5\times10^5$ steps.

To model the HCP networks, we have used two different bonded interaction $u_{\rm b}(r)$: namely the
finitely extensible nonlinear elastic (FENE)~\cite{kremer1990dynamics} and the quartic potential~\cite{stevens2001interfacial,TingeRobinsQuartic}.

\subsection{Network cure}

\begin{figure}[ptb]
\includegraphics[width=0.49\textwidth,angle=0]{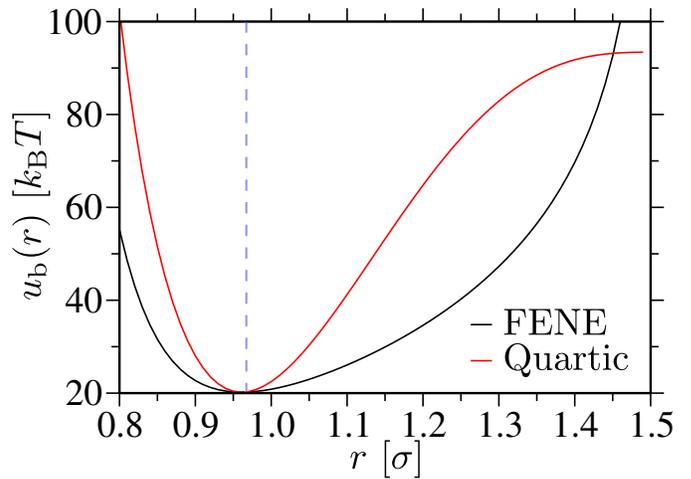}
	\caption{Interaction potentials between the bonded monomers $u_{\rm b}(r)$ as a function of inter monomer distance $r$. 
	The Quartic parameters are taken from the literature~\cite{TingeRobinsQuartic}. The equilibrium bond length for these models is 
	around $\ell_{\rm b} = 0.97\sigma$ represented by the vertical line. 
\label{fig:pot}}
\end{figure}

During the network formation FENE is used, where a bond between two monomers 
is defined by the combination of repulsive 6-12 LJ potential,
\begin{equation}
u_{\rm b}(r) = 4\epsilon \left[\left(\frac {\sigma} {r}\right)^{12} - \left(\frac {\sigma}{r}\right)^6 
	+ \frac {1}{4}\right]~ {\rm for}~ r < 2^{1/6}\sigma_{\rm b},
\end{equation}
and the FENE potential,
\begin{equation}
\label{eq:fene}
	u_{\rm FENE}(r) = -\frac {1} {2} k R_{\circ}^2 \ln \left[1 - \left(\frac {r}{R_{\circ}}\right)^2\right].
\end{equation}
Here, $k = 30 k_{\rm B}T/\sigma^2$ and $R_{\circ} = 1.5\sigma$.
This gives a typical bond length of $\ell_{\rm b} \simeq 0.97\sigma$ ~\cite{kremer1990dynamics}, 
see the black curve in Fig.~\ref{fig:pot}. 

\begin{figure}[ptb]
\includegraphics[width=0.49\textwidth,angle=0]{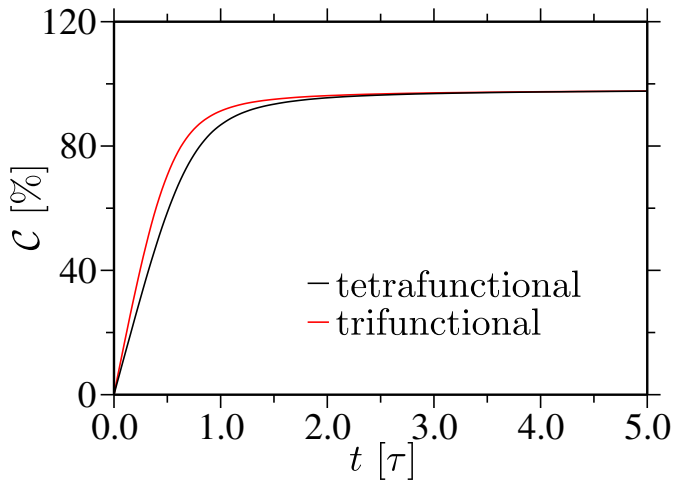}
	\caption{Curing percentages $C$ during the formation of total number of bonds $\mathcal{N}_{\rm b}$ with time $t$ for 
	two different network functionalities $n$. The data is shown for the tri-functional (i.e., $n=3$) and the tetra-functional (i.e., $n=4$) systems. 
	Network cure is performed via FENE bonds, see Eq.~\ref{eq:fene}. Note that for the representation purpose we have only shown 
	results for the initial time of $5\tau$, while the total simulation time is $5 \times 10^3 \tau$.
	\label{fig:cure}}
\end{figure}

Because the indentation is performed along the $z-$direction, the starting configuration (before curing) consists of
a homogeneous sample of LJ particles at $\rho_{\rm m}$ confined between two repulsive walls along the 
$z-$direction. The periodic boundary conditions are employed in the $x$ and the $y$ directions. 
The bonds are allowed to form between the monomers using a protocol proposed in the earlier works of 
some of us~\cite{DMPRM21, MukherjiPRE2009}. Within this protocol, the bonds are randomly 
formed between two monomers when: 1) two particles are closer than $1.1\sigma$
distance, 2) monomers have not formed the maximum number of possible bonds limited by $n$, 
and 3) a random number between zero and one is less than the bond forming probability of 0.05.
The network curing is performed for $t_{\rm cure}=5\times 10^3\tau$
during the canonical simulation. Here, $t_{\rm cure}$ is sufficient to attain close 
to 99\% cure, see Fig.~\ref{fig:cure}.
This procedure also ensures a rather homogeneous bond formations within the sample, see the Supplementary Fig.~S1~\cite{epaps}.

After the network curing stage, systems are equilibrated for $t_{\rm NPT} = 2.55 \times 10^3 \tau$ at the zero pressure. Here, pressure is 
employed using Nose-Hoover barostat with pressure damping parameter  $\gamma_{\rm p} = 0.5 \tau$.

\subsection{Computational indentation}

For the indentation simulations, we have used a quartic potential~\cite{TingeRobinsQuartic},
\begin{equation}
\label{eq:quart}
u_{\rm quartic}(r)=k{\tilde r}^2({\tilde r}-B_1)({\tilde r}-B_2)+u_{\circ}. 
\end{equation}
Here, $k = 103.2 {k_{\rm B}T}/{\sigma^4}$, ${\tilde r} = r-R_c$, $R_c = 1.495 \sigma$, $B_1 = -1.001 \sigma$, $B_2 = 1.426 \sigma$,
and $u_\circ = 44.8 k_{\rm B}T$. This potential ensures bond breaking if the distance between
two bonded monomers is larger than $R_c$.

An implicit spherical indenter of radius $R$ is used. The force acting on the indenter is obtained by 
summing the force contributions of the first shell surrounding monomers. Here, the individual force is calculated
using the derivative of $u_{\rm indenter}(r)=K(r-R)^2$, where $K=100~k_{\rm B}T/\sigma^2$ 
is the force constant and $r$ is the distance from the atom to the center of tSuchhe indenter. 
The indentation is performed at a constant velocity $v = 0.005 \sigma/\tau$
and in the microcanonical ensemble. Two different indenter with $R = 5.0\sigma$ and 15.0$\sigma$ are chosen for this work. 

We note in passing that we have also conducted simulations 
at $v = 0.05\sigma/\tau$. 
The results showed no noticeable difference for the tetrafunctional sample, while a significant difference was 
observed for the trifunctional system (data not shown). Therefore, we have chosen $v = 0.005\sigma/\tau$ below which
the mechanical response remain invariant irrespective of the network functionality. To test the reproducibility of the data, we have performed three sets of simulations for $R = 5.0\sigma$ and for both functionalities. The data is shown in the Supplementary Section S6~\cite{epaps}.

\section{Results and discussion}
\label{sec:res}

\subsection{Force response upon indentation}

\begin{figure}[ptb]
\includegraphics[width=0.49\textwidth,angle=0]{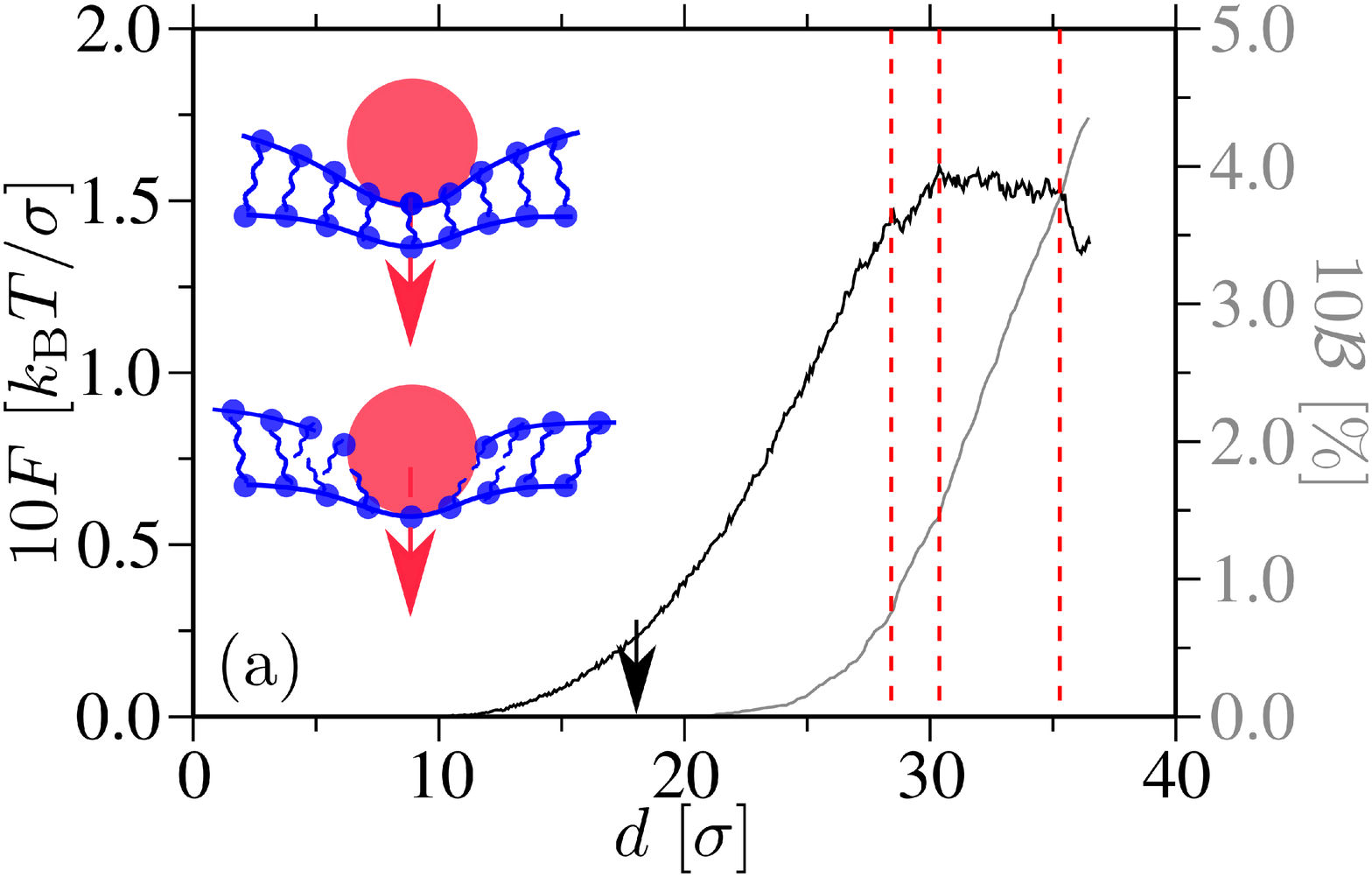}
\includegraphics[width=0.49\textwidth,angle=0]{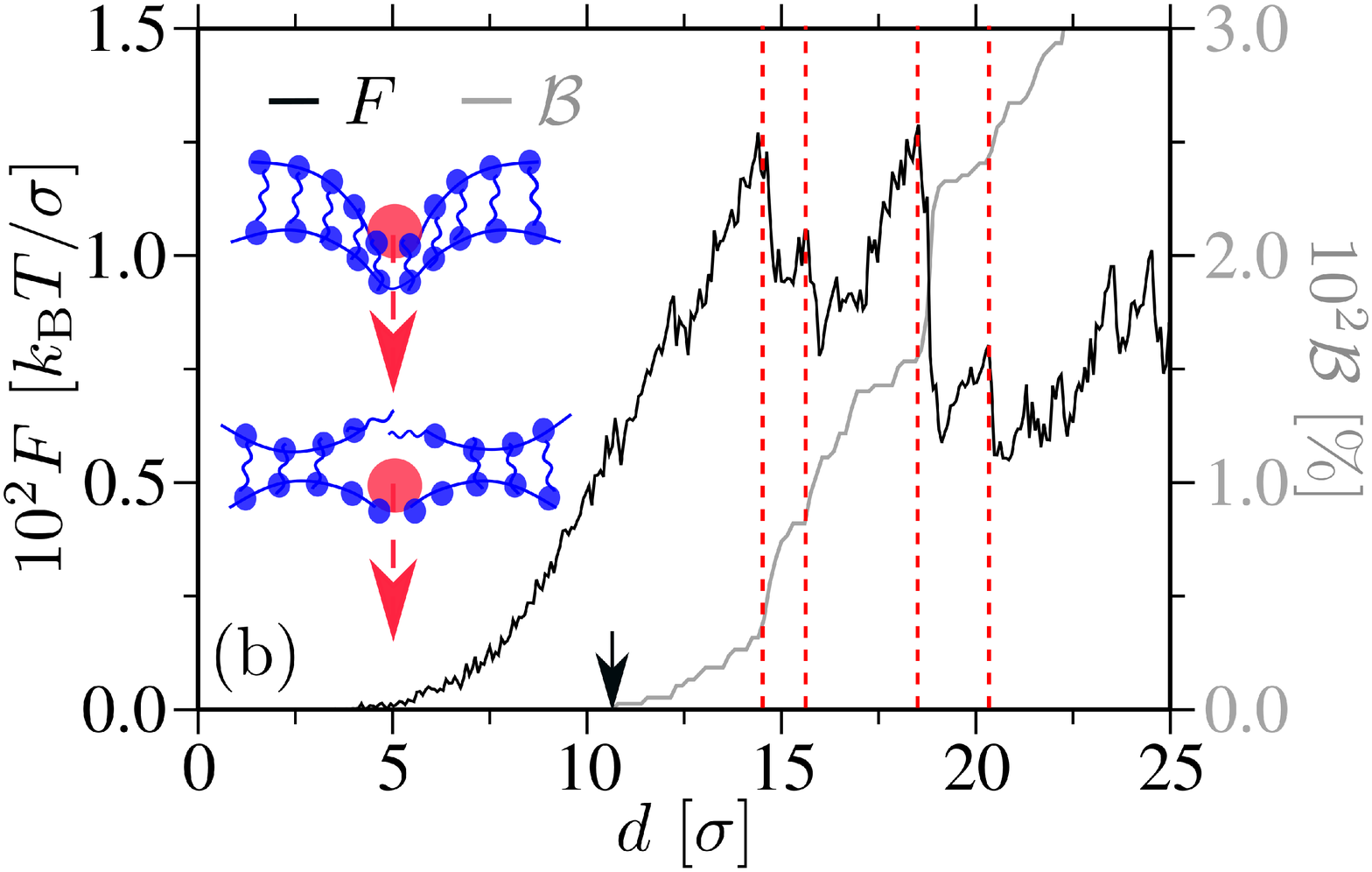}
	\caption{Force $F$ and the percentage of broken bonds $\mathcal{B}$ as a function of the indentation depth $d$. 
	Parts (a) and (b) show the data for the indenter radius $R = 15.0\sigma$ and $5.0\sigma$, respectively.
	The color codes of the data sets are consistent with their corresponding $y-$axes.
	Data is shown for an indentation velocity of $0.005\sigma/\tau$ and for a tetra-functional network. 
	Vertical lines highlight the major drops in $F$ and the corresponding $\mathcal{B}$. Initial depletion zones 
	below $d < 10.0\sigma$ (in part a) and $d < 5.0\sigma$ (in part b) are because the lowest part of the indentation 
	tips are certain distance away from the network surface. Arrow indicates at the $d$ values where the 
	bond breaking starts. The insets show different pictorial representations of the indentation tips entering 
	the sample at different $d$. The top and the bottom panels show the cases just before and after the bond breaking
	in the samples, respectively.
	\label{fig:fd1}}
\end{figure}

We start by discussing the mechanical response of a tetrafunctional HCP network. 
Fig.~\ref{fig:fd1} shows the typical force $F$ versus indentation $d$ behavior for two different $R$.
Other than the generic behavior, i.e., the increase in $F$ with increasing $d$, one interesting feature can be seen 
is the sudden force drop $\Delta F$ at various $d$. Note that we define $\Delta F$ only when its magnitude
is larger than the percentage of error fluctuation calculated in the elastic regime, i.e., below bond breaking indicated by the black arrows in Fig~\ref{fig:fd1}.
Such force drop is a well known phenomenon in the mechanical response of the glassy materials, where 
the atomic rearrangements during deformation can lead to such force drops~\cite{falk1998dynamics}.
 This phenomenon is commonly referred to as the {\it avalanche}. 
In our study, however, we are dealing with a rather rigid cross-linked network that has 
microscopically different molecular connectivity in comparison to the common glasses.

What causes such avalanche-like deformation in HCP? To investigate this issue, we have calculated the 
percentage of bond breaking $\mathcal{B}$ as a function of $d$. 
It can be appreciated that $\Delta F$ (see the black data sets in Fig. \ref{fig:fd1}) is directly related to the 
large number of broken bond (see the light grey data sets in Fig. \ref{fig:fd1}). This behavior is quite expected 
because the strongly interacting bonded monomers can significantly resist the
deformation. When the bonds break, they induce large force drops in HCPs. 
In the insets of Fig.~\ref{fig:fd1} we show the possible pictorial representations of the cases just before and 
after the bond breaking in these samples. Such avalanches, induced by the bond breaking, were also observed in a combined experimental and simulation study on soft polymer networks \cite{boots2022quantifying}.

Fig. \ref{fig:fd1} also reveals that $\Delta F$ is more prominent for $R = 5.0\sigma$ in comparison to $R = 15.0\sigma$.
This behavior is not surprising given that the smaller indenter can delicately monitor the bond breaking that
occurs at the monomer level, while the larger indenter tip can only monitor a larger group of monomers that
on average contribute to $F$ and $\mathcal B$. We also note in passing that the standard tensile deformation 
can only investigate the bulk mechanical behavior~\cite{MukherjiPRE2008}, while the indentation simulations can 
reveal the small-scale microscopic structural details because of the localized deformation, which is 
also intrinsically related to $R$. Additionally, the relatively larger $F$ values for $R=15.0\sigma$ 
is because this indenter tip has on average a larger number of monomers that are directly in contact with it.

\begin{figure}[ptb]
\includegraphics[width=0.49\textwidth,angle=0]{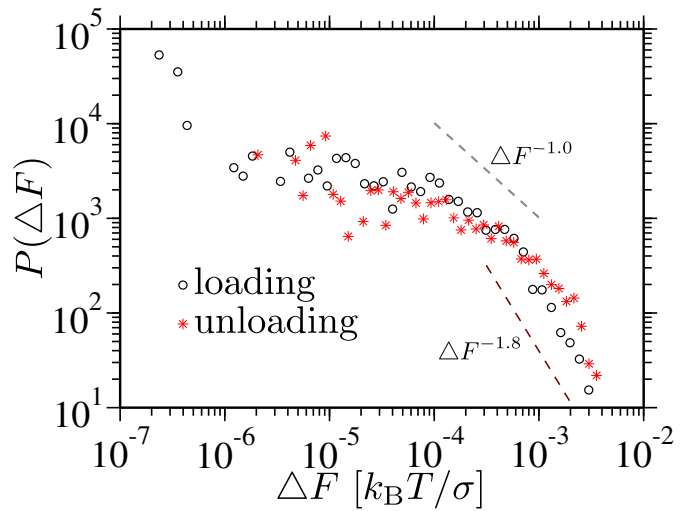}
	\caption{Distribution of the force drops $P(\Delta F)$ obtained from Fig.~\ref{fig:fd1}(b). $\Delta F$ is calculated at the onset of bond breaking in the tetrafunctional networks. The data are shown for an indenter radius $R = 5\sigma$. The lines are power law fits, i.e., $\triangle F^{-1.0}$ (for $10^{-4} < \triangle F < 10^{-3}$) and $\triangle F^{-1.8}$ (for $\triangle F > 10^{-3}$). We have also added the data corresponding to the unloading cycle. The data are averaged over three different simulation runs, as described in the Supplementary Section S6~\cite{epaps}.
	\label{fig:fd2}}
\end{figure}

As discussed in the preceding paragraphs, the observed trends in $\Delta F$ is reminiscent of the mechanics of amorphous materials. 
It would therefore be interesting to see whether the distribution of $\Delta F$ follows a scaling law as it is the case for sheared amorphous systems~\cite{SalernoRobinsaPRE2013}. For this purpose, we have also calculated distribution of $\Delta F$ for our systems $P(\Delta F)$.
The corresponding data is shown in Fig.~\ref{fig:fd2}. It can be appreciated that the data shows three regimes: a plateau for $\triangle F < 10^{-4}$, $P(\Delta F) \propto {\Delta F}^{-1.0}$ for a narrow region $10^{-4} < \triangle F < 10^{-3}$, and finally $P(\Delta F) \propto {\Delta F}^{-1.8}$ for $\triangle F > 10^{-3}$. For the elastic-plastic deformation of glassy systems, most of the studies have focused on the distribution of avalanches in the stationary regime. In these cases, three-dimensional simulations have reported the exponents within the range of $1.2-1.3$~\cite{SalernoRobinsaPRE2013, ozawa2018random}. 
However, only a few works have dealt with the transient regime and results obtained so far suggest either a similar value of the avalanche exponent as in the stationary regime~\cite{ozawa2018random} or smaller~\cite{ruscher2021avalanches}.
In this context, the gel phase of HCP systems investigated here are microscopically different from the traditional mono-atomic glass formers. Therefore, we can only point toward the relatively close value of the avalanche exponent without attempting to state precisely why this description still holds in our case. Moreover, the close resemblance of the mechanical behavior observed between two
microscopically distinct systems may direct at a more generic mechanical picture of the jammed systems. 
We also note in passing that the goal of this work is not to study the scaling laws of $P(\Delta F)$, 
rather to only study any possible close resemblance with the avalanche-like behavior.
Therefore, we abstain from going into further details on this aspect.

\subsection{Mechanics during loading and unloading}

In the typical elastomers, microgels, hydrogels and/or in polymers, the deformation is 
usually visco-elastic in nature upon small deformations. However, mechanical indentation in HCPs 
also induces a significant amount of bonds breaking, thus is expected to follow the 
elastic-plastic deformation. In this context, it has been previously shown that the depth 
sensing of the materials can be investigated within the loading-unloading setup~\cite{oliver2004measurement}.
Furthermore, the hardness, the stiffness and the effective elastic modulus of a sample can be 
readily calculated. Therefore, in this section we investigate HCPs under cyclic loading.

\begin{figure*}[ptb]
\includegraphics[width=0.49\textwidth,angle=0]{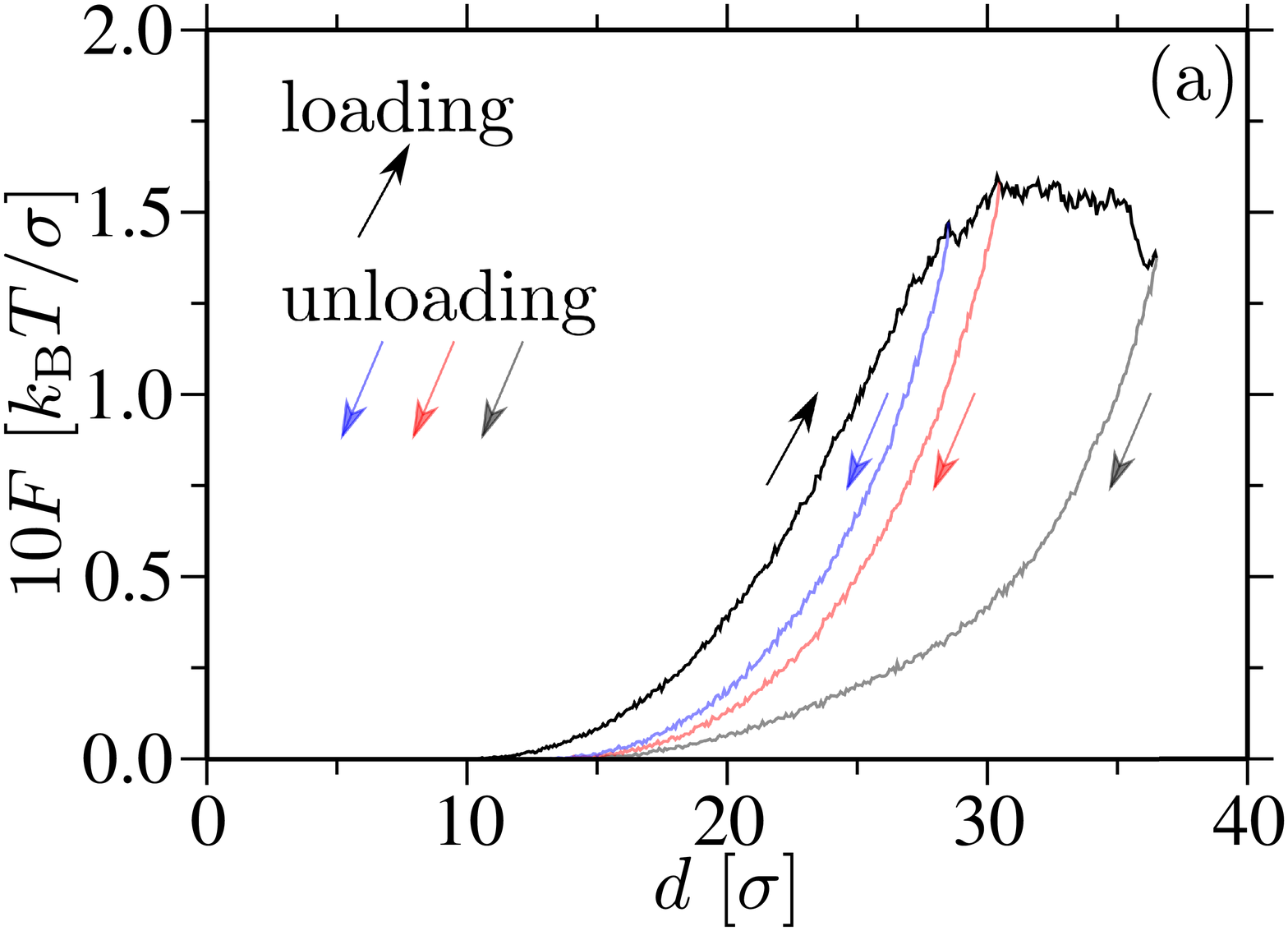}
\includegraphics[width=0.49\textwidth,angle=0]{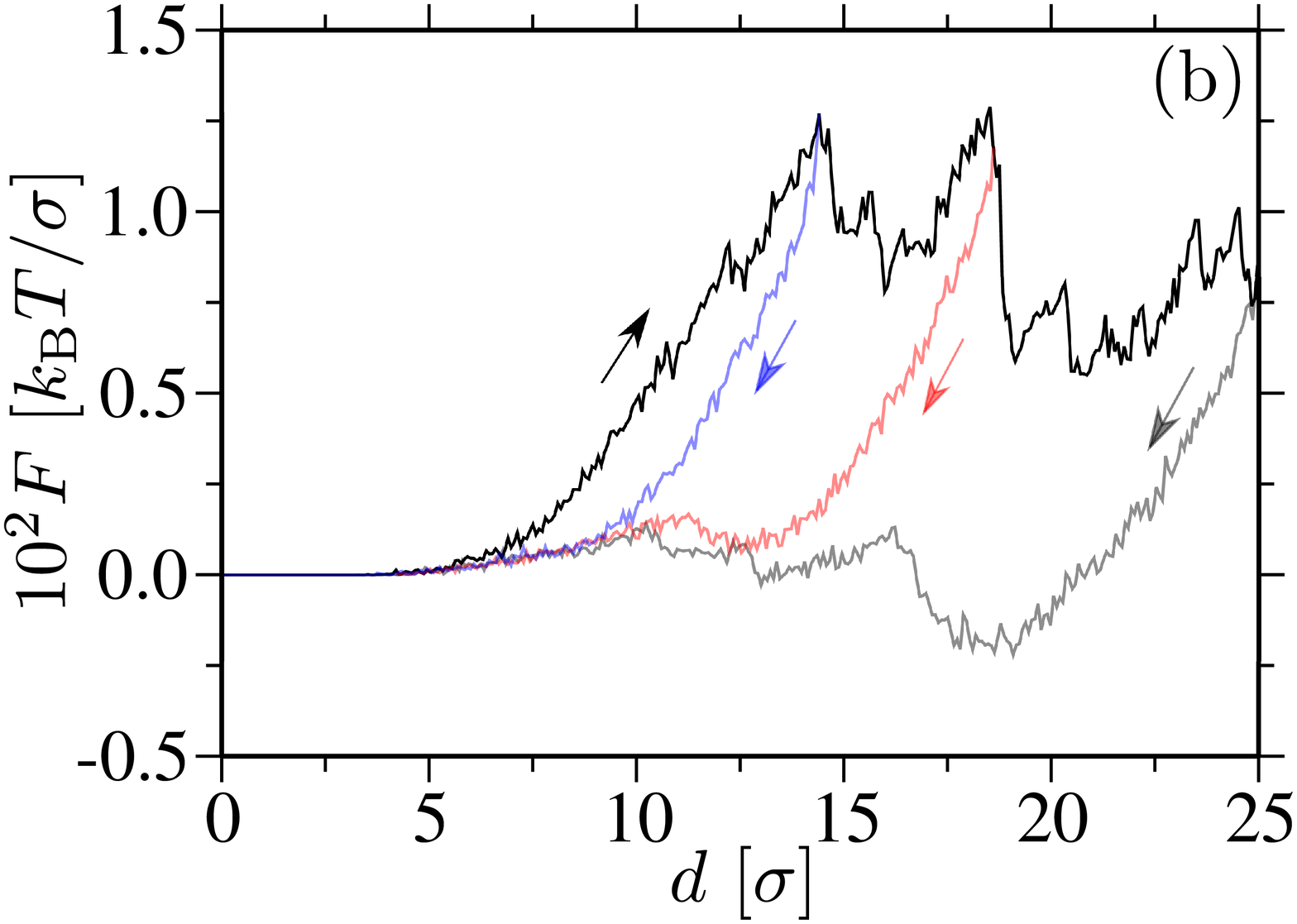}
	\caption{Same as Fig.~\ref{fig:fd1}, however, during both loading and unloading cycles.
	Arrows indicate at the corresponding loading and unloading curves. Parts (a) and (b) 
	show the data for the indenter radius $R = 15.0\sigma$ and $R = 5.0\sigma$, respectively.
	For clarity of presentation, we have only shown four data sets in part (a). More
	data is presented in the Supplementary Fig.~S2 \cite{epaps}.
	\label{fig:fd3}}
\end{figure*}

Fig.~\ref{fig:fd3} shows the mechanics of HCPs during the loading and the unloading cycles. 
The significant hysteresis indicate at a large plastic deformation in the samples, resulting from 
a large number of the broken bonds. Furthermore, the strong undershooting, i.e., $F < 0$ for $R = 5.0\sigma$,
in Fig.~\ref{fig:fd3}(b) is because an unloading cycle pulls certain number of monomers
that induces a negative pressure in the samples. This behavior is in good agreement with a set of recent experiments on hydrogels, where large force drops during the loading cycles and strong hysteresis during the unloading cycles were observed~\cite{muthukumar2022cone}. It should also be highlighted that the most severe for effects of hysteresis is observed for $d \geq 2R$. We will come back to this point at a later stage of this draft.

\begin{table*}[ptb]
        \caption{A table listing the observables for the calculation of effective elastic modulus $E_{\rm eff}$
        in Eq.~\ref{eq:eff}.
        Data is shown for the tetrafunctional networks and for two different indenter radius $R$. $d_{\rm max}$ and 
        $F_{\rm max}$ are the maximum displacement and force just before the unloading, respectively.
        The slope $S$ is calculated from the initial linear regime during unloading. $d_c$ 
        is the contact depth between the spherical indenter and the sample and $\mathcal A$ is the contact area.
        Note that for the calculation of $d_{\rm max}$, the initial depletion zones in Fig.~\ref{fig:fd3} 
        are subtracted.}
\begin{center}
       \begin{tabular}{|c|c|c|c|c|c|c|c|c|c|c|c|}
\hline
{\color{white}-}&&&&&&\\
$~~~R~[\sigma]~~~$ & $~d_{\rm max}~[\sigma]~$ &   $~F_{\rm max}~[k_{\rm B}T/\sigma]~$ & $~S = \frac {{\rm d}F}{{\rm d}d}~[k_{\rm B}T/\sigma^2]~$  & $~~~d_c~[\sigma]~~~$ & ~~~${\mathcal A}~[\sigma^2]~~~$ & $~10^{2}E_{\rm eff}~[k_{\rm B}T/\sigma^3]~$\\
{\color{white}-}&&&&&&\\
\hline
\hline
\hline
{\color{white}-}&&&&&&\\
15.0 &  4.1 &  0.006  &   0.0029    &   2.5    &   215.98   &    17.49\\
{\color{white}-}&&&&&&\\
& 9.0 &  0.038  &   0.0087    &   6.5   &    479.88   &  35.20\\
{\color{white}-}&&&&&&\\
& 15.8&  0.111  &   0.0194    &   11.5    &   668.37   &  66.50\\
{\color{white}-}&&&&&&\\
& 18.5 &  0.147  &   0.0226    &   13.6    &    700.70   &  75.66\\
{\color{white}-}&&&&&&\\
& 20.5 &  0.158  &  0.0234    &   15.4   &     706.36   &      78.03\\
{\color{white}-}&&&&&&\\
& 26.5 &  0.137  &   0.0204    &   21.5    &   574.13   &  75.45\\
{\color{white}-}&&&&&&\\
\hline  
\hline
{\color{white}-}&&&&&&\\
5.0 & 10.4 &  0.013  &  0.0038 &7.9&52.12&46.65\\
{\color{white}-}&&&&&&\\
& 14.6 &  0.013  &  0.0025 &10.7&--&--\\
{\color{white}-}&&&&&&\\
&21.0 &  0.008 &  0.0021 & 18.1&--&--\\
{\color{white}-}&&&&&&\\
\hline
\end{tabular}  \label{tab:eff}
\end{center}
\end{table*}

Having shown the mechanics during the loading and the unloading cycles, we will now investigate if 
the relevant system properties can be extracted from the data in Fig.~\ref{fig:fd3}. 
In this context, there is a theory that connects the mechanics under the unloading cycles 
to the material stiffness via the ``so called" effective elastic modulus $E_{\rm eff}$~\cite{oliver2004measurement},
\begin{equation}
\label{eq:eff}
    E_{\rm eff}  = \frac {S}{2} \sqrt{\frac {\pi}{\mathcal A}}.
\end{equation}
Here, $S = {{\rm d}F}/{{\rm d}d}$ is calculated from the initial displacements during the unloading cycles.
The contact area for a spherical indenter can be estimated using ${\mathcal A} = 2 \pi R d_c - \pi d_c^2$ and 
the contact depth between the indenter and the sample is $d_c = d_{\rm max} - 0.75 {F_{\rm max}}/S$.
We note in passing that this theory was initially developed for the Berkovich contact~\cite{oliver2004measurement}. It is widely used in typical micro- and nano-indentation experiments. A more general treatment is discussed in that dealt with different indenter shapes~\cite{oliver2004measurement}. 

Using Eq.~\ref{eq:eff} we have calculated $E_{\rm eff}$ during the unloading at different 
indentation depths and for both $R$. The data is compiled in Table~\ref{tab:eff}.
For $R = 15.0\sigma$, it can be appreciated that $E_{\rm eff}$ increases with increasing $d_{\rm max}$, 
before reaching a plateau at $d_{\rm max} > R$. This length scale is also consistent with the $d$ value at which the force plateaus in Fig.~\ref{fig:fd3}

The stiffening upon deformation is reminiscent of the strain-hardening behavior in polymeric materials~\cite{MukherjiPRE2009, hoy2007strain}.
 In this context, it has been previously shown that a 
tetrafunctional HCP can strain-harden under tensile deformation \cite{MukherjiPRE2008}. There it was argued that the network curing protocol induces a significant amount of free volume
within the cured samples. These free volumes usually collapse to form rather large protovoids centers, where
the monomers around the periphery of a protovoids can statistically form bonds pointing away from each other.
During the tensile deformation, these protovoids open up by disrupting the van der Waals (vdW) contacts and 
thus may be a possible cause for the strain hardening. In contrast to the tensile testing, indentation
compresses a sample. Such compression induced hardening results from two effects: 
(1) The small-scale tension buildup when the certain number of bonds are pulled taut, while some other are compressed. 
(2) The two boundaries of a protovoid interacting via vdW forces can plastically slide past each other.
These two combined effects lead to the major contributions to the increased $E_{\rm eff}$, that we name {\it indentation hardening}. 

Table~\ref{eq:eff} also shows that for $d_{\rm max} > 2R$ and $R=5.0\sigma$, 
$E_{\rm eff}$ is not defined. This is particularly because Eq.~\ref{eq:eff} only holds for the small 
relative indentations, i.e., when $d_c \leq 2R$.

The theoretical treatment described above does not account for the dynamics, such as the systems 
exhibiting visco-elastic deformation. Here, one may expect the
force relaxation to be significantly faster than the structural relaxation. On the contrary, a HCP network is microscopically different, where the system shows a 
elastic-plastic deformation via bond breaking during the indentation, see Fig.~\ref{fig:fd3}. 
Furthermore, we find that the relaxation of $F$ is significantly slower than the structural relaxation, see the Supplementary Fig.~S3 \cite{epaps}. 
We also wish to highlight that the HCP networks investigated here are in their gel phase where the high frequency elastic contacts due to the chain connectivity in the network 
architectures play the crucial role in restoring the network structures. The mode of this temporal relaxation is shown in
a supplementary simulation movie.

\subsection{Effect of network functionality}

The network functionality $n$ is also expected to significantly impact the force response. 
For this purpose, here we show the system with $R=5.0\sigma$ where the most prominent force drops are observed, see Fig.~\ref{fig:fd1}(b), while the data for $R=15.0\sigma$ is presented in the Supplementary Fig.~S4.
\begin{figure}[ptb]
\includegraphics[width=0.49\textwidth,angle=0]{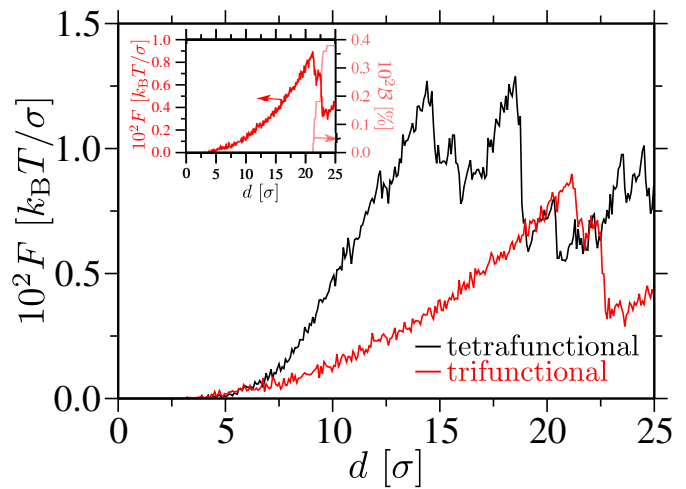}
	\caption{Same as Fig.~\ref{fig:fd1}(b), however, for two different
	network functionalities. The inset also includes the percentage of
	bond breaking $\mathcal B$ for the trifunctional network. For this
	network, bond breaking starts at around a depth of $d \simeq 21.0\sigma$.
	Data is shown for an indenter radius $R=5.0\sigma$.
	\label{fig:fd4}}
\end{figure}
Fig.~\ref{fig:fd4} shows the effect of $n$ on the mechanical response. It can be appreciated that the 
trifunctional system can withstand almost twice larger indentation depth before the bond breaking happens in
a sample. For example, bond breaking starts at $d\simeq 11.0\sigma$ for a tetrafunctional sample (see Fig.~\ref{fig:fd1}a), 
while it is about $d \simeq 21.0\sigma$ for a trifunctional system (see the inset in Fig.~\ref{fig:fd4}). 
This is particularly because the lower bond density of a trifunctional sample, in comparison to a tetrafunctional sample,
can withstand a larger elastic deformation. Furthermore, a trifunctional system is twice as ductile as a 
tetrafunctional sample, while the maximum force these samples can withstand also reduces by about a factor of two for the former 
(see the main panel in Fig.~\ref{fig:fd4}). These results are reasonably consistent with the HCPs under
the tensile deformations~\cite{MukherjiPRE2009}.

\subsection{Effect of network curing percentage}

\begin{figure}[ptb]
\includegraphics[width=0.49\textwidth,angle=0]{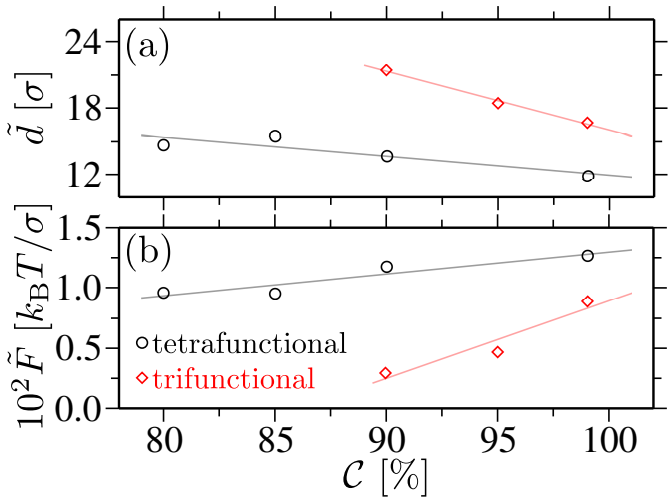}
	\caption{The first maximum at the force drop $\tilde F$ and corresponding depth $\tilde d$ 
	as a function of the percentage of the network cure $\mathcal C$. The data is shown for
	two different network functionalities and for an indenter radius $R=5.0\sigma$. 
	Note that for the calculation of $\tilde d$, the initial depletion zones in the 
	force-extension curves are subtracted.
	\label{fig:fd5}}
\end{figure}

One of the most important features of the HCP networks is that their mechanics can be tuned almost {\it at will}
by tuning the amount of cure in the sample. Therefore, in this section we aim to investigate the effect of curing
on the mechanics of the samples. For this purpose, we have calculated the force-indentation behavior for
different percentage of cure $\mathcal C$, see the Supplementary Fig.~S5. 
From these data sets, we extract 
the force at the first maximum drop $\tilde F$ and the corresponding indentation depth $\tilde d$.
Fig.~\ref{fig:fd5} shows the variation of $\tilde F$ and $\tilde d$ with the change in $\mathcal C$.
The data clearly indicate a reasonable linear variation that can serve as a guiding tool for 
the possible synthesis of epoxies with tunable mechanical response.

\section{Conclusion and Discussion}
\label{sec:disc}

Using large scale molecular dynamics simulations of a (mesoscopic) generic model, we investigate the mechanical
response of a set of highly cross-linked polymer (HCP) networks. For this purpose, we have used the
mechanical indentation of different indenter sizes. The use of such method has many advantages
over the routinely used tensile deformations. For example, the deformations at the monomer level
can be monitored because the individual bond breaking and network (re-)arrangements are directly
accessible. Contrary to the visco-elastic deformations in the standard polymers, the deformations in the 
HCPs are of elastic-plastic nature. An analysis based on a simple theory of elastic-plastic
deformation~\cite{oliver2004measurement}, suggests that the HCP networks harden upon indentation.
This is because the bonds that are pulled or compressed taut and the disruption of 
the protovoids (collapsed free volume area) that originate within the sample 
during the network cure~\cite{DMPRM21}.

The results also indicate at a very intriguing force-drop $\Delta F$ like mechanical response, reminiscent of the avalanche in the glassy materials.
Such $\Delta F$ behavior in HCPs are related to the instantaneous bond breaking, small-scale monomeric rearrangements,
and localized force response within the samples. We further show that the distribution of $\Delta F$ follows a scaling 
law decay with an exponent close to the theoretical predictions for the avalanches in the standard glasses.

The results presented in this work is based on a generic molecular dynamics approach
that combines a broad range of chemical systems within one physical framework. While 
such models are extremely useful in studying the trends observed in the chemical 
specific systems, they do not provide any quantitative agreement with the experiments.
Also note that there are generic models that are specifically tuned to reproduce certain 
quantities and their respective behavior~\cite{DM20macromolwithRobin}.
Therefore, it is important to discuss the possible chemical systems that represent
the HCP networks discussed here. In this context, in a generic model, one bead corresponds 
to a certain number of atomistic monomers ~\cite{kremer1990dynamics, stevens2001interfacial}, while
the degree of such coarse-graining is controlled by the bond and bending stiffnesses.

The experimental synthesis of HCPs, such as the epoxies, are performed with different underlying chemistry. 
For example, the most commonly used systems include, but are not limited to: the amine-cured epoxies ~\cite{sharifi2014toughened} and 
the phenylenediamines-based cross-linking of a network structure ~\cite{lv2020effect}. These systems 
are rather rigid network materials that exhibit extraordinary mechanical ~\cite{sharifi2014toughened, MukherjiPRE2008} and 
thermal properties ~\cite{DMPRM21, lv2021effect}. Furthermore, more complex structures are also synthesized that 
use the alkane chains for cross-linking \cite{lv2021effect}. The longer the length of the alkanes, the softer the network 
at a given temperature ~\cite{lv2021effect}. In this context, the quartic potential parameters used 
in this work closely mimics the amine-based epoxies, while the FENE bonds mimic the relatively softer bonded 
interactions induced by the alkane-based linkers. A more detailed discussion on the atomistic to
generic mapping schemes is presented in Ref.~\cite{DMPRM21}.

It is also important to discuss that the mechanical testing on weakly cross-linked systems, such as the 
PAM-based hydrogels ~\cite {muser2019modeling}, surface-grafted gels ~\cite{singh2018combined}, and PNIPAM-based microgels \cite{backes2017combined} 
often lead to unexpected behavior. One of the interesting aspects of these systems is that they exhibit
depth dependent force response, i.e., the stiffness of the samples change with the depth. 
This is practically because of the synthesis protocols that unavoidably introduce a greater degree 
of bond density at the core and decreases radially outward from the centers. 
In this context, while our trifunctional system may mimic such weakly cross-linked networks, the bond density in our systems are rather homogeneous across the sample, see the Supplementary Fig.~S1 \cite{epaps}. Therefore, a more careful modelling will require an additional control parameter of the non-homogeneous
bond density within the samples that can serve as a guiding tool for the materials design. A detailed discussion on this aspect will be presented elsewhere.\\

\noindent{\bf Note added during the revision:} After submission of our work on 7th February 2022, two important related experimental works were published that we cite in this revised draft~\cite{boots2022quantifying,muthukumar2022cone}.


\section{Acknowledgement}
D.M. thanks Martin M\"user for useful discussions. This research was undertaken thanks, in part, to the 
Canada First Research Excellence Fund (CFREF), Quantum Materials and Future Technologies Program. MKS thanks Science and Engineering Research Board (SERB), India for financial support provided under the Start-up Research Grant (SRG) scheme (grant number: SRG/2020/000938).


\bibliographystyle{ieeetr}
\bibliography{reference.bib}


\end{document}